\documentclass[pdflatex,sn-mathphys-num]{sn-jnl}% Math and Physical Sciences Numbered Reference Style
\usepackage{graphicx}%
\usepackage{multirow}%
\usepackage{amsmath,amssymb,amsfonts}%
\usepackage{amsthm}%
\usepackage{mathrsfs}%
\usepackage[title]{appendix}%
\usepackage{xcolor}%
\usepackage{textcomp}%
\usepackage{manyfoot}%
\usepackage{booktabs}%
\usepackage{algorithm}%
\usepackage{algorithmicx}%
\usepackage{algpseudocode}%
\usepackage{listings}%
\newcommand{\be}{\begin{equation}}
\newcommand{\ee}{\end{equation}}
\newcommand{\bea}{\begin{eqnarray}}
\newcommand{\eea}{\end{eqnarray}}
\theoremstyle{thmstyleone}%

\theoremstyle{thmstyletwo}%
\theoremstyle{thmstylethree}%
\begin{document}

\title[Article Title]{Integrating Heisenberg uncertainty and maximum entropy principles verses statistical approach to nucleon structure}

\author*[1]{ \sur{A.~Mirjalili}}\email{a.mirjalili@yazd.ac.ir}

\author[1]{ \sur{S.~Khosravi Kia}}\email{simsim937@yahoo.com}

\author[2,3]{ \sur{S.~Atashbar Tehrani}}\email{atashbart3@gmail.com}

\affil[1]{\orgdiv{Physics Department}, \orgname{Yazd University}, \orgaddress{\street{} \city{Yazd},  \country{Iran}}}

\affil[2]{\orgdiv{School of Particles and Accelerators}, \orgname{Institute for Research in Fundamental Sciences (IPM)}, \orgaddress{\street{} \city{Tehran}, \postcode{19395-5531} \state{}, \country{Iran}}}

\affil[3]{\orgdiv{Department of Physics}, \orgname{ Faculty of Nano and Bio Science and Technology, Persian Gulf University}, \orgaddress{\street{} \city{Bushehr}, \postcode{75169},  \country{Iran}}}

\abstract{There are many methods to calculate the parton distributions. In this work, we are trying to determine the unknown parameters of valence quark  and gluon densities, taking into account some constraints simultaneously. For this purpose, we consider two different approaches. At first approach, a parameterized form for the valence densities are assumed. Using the constraints imposed by the  Heisenberg uncertainty and maximum entropy principles and also considering the quark number and momentum sum rules, the unknown parameters of the parton distributions are determined. In second  approach, we attribute a statistical distributions to quark, anti-quarks and gluon densities which are including temperature T, volume V and chemical potential, $\mu$, as thermodynamic parameters. In this case, using only the maximum entropy principle and the quark number together with momentum sum rules, the parton distributions in terms of thermodynamic parameters are extracted. The results for valence quark  densities in both approaches and additionally the gluon density in second approach  are in {satisfactory match} with results from some parametrization models. By evolving these densities to the desired energy scales, we are able to calculate the nucleon $xF_3$ structure function and also the ratio of valence quark densities.}




\maketitle
\section{Introduction}\label{sec:intro}
Identification of parton distributions and the resulted nucleon structure functions are still some topics of interest in theory of Quantum  Chromodynamics (QCD). There are different methods to determine these functions.  The prevalent method depends on to  select  specific functions  in terms of  x-Bjorken variable that contains several unknown  parameters. The fitting of functions will be achieved by comparing the evolved parton distributions with available experimental data which are expected to give us potentially physical results. These functions can be used to calculate the hadron cross section and also to determine the nucleon structure  function.
The initial parton distributions at low energy scale, $Q_0$, as the inputs of the calculations, are in fact the  non-perturbative part of concerned computations and we intend to obtain them in two different approaches. In the naive parton model, the statistical quantum correlations which exists between the initial and final states of parton distributions, are ignored. In statistical approach these correlations are considered by assuming Fermi-Dirac distributions for quarks and Boose-Einstein distributions for gluons \cite{bou}.\\

In this paper, we try  at first to determine parton distribution functions, using  Heisenberg uncertainty and  maximum entropy principles together with quark number and momentum sum rules. In the second step, to achieve parton distributions, we assume a statistical distribution for partons and use only  the maximum entropy principle together with the referred sum rules. {Based on this foundation, the unknown quark density parameters are calculated, allowing for the related computations of the evolved parton densities and nucleon structure functions to be performed.}\\

The organization of this paper is as it follows. In Sec.\ref{sec:2} we give a brief review on the current sum rules in parton model. How to employ the Heisenberg uncertainty principle in parton model is discussed in Sec.\ref{sec:3}. We deal with the principle of  maximum entropy,  considering the  parton densities in Sec.\ref{sec:4}. Extracting the unknown parameters of parton densities at initial energy scale ,$Q_0$, taking into account the  Heisenberg uncertainty and maximum entropy principles together with the available sum rules are argued in  Sec.\ref{sec:5}. Evolution the parton densities to determine them in high energy scale and also to calculate the $xF_3$ nucleon structure function are performed and the results are also presented in Sec.\ref{sec:6}. { As prerequisites for employing our second approach, the fundamental concepts of the statistical method for parton densities are initially presented in Sec.\ref{sec:7}.}
This subject is followed, considering the light cone coordinates in Sec.\ref{sec:8}. Numerical results for unknown parameters of parton densities in statistical approach, based on the current sum rules and only the principle of maximum entropy is presented in Sec.\ref{sec:9}. Phenomenological achievements of statistical approach are indicated in Sec.\ref{sec:10} and  finally we give our discussion and conclusion in Sec.\ref{sec:11}

\section{Constraints: Quark number and momentum sum rules}\label{sec:2}

In describing the hadron spectroscopy, quark model has an essential rule. In fact this model makes a possibility to reveal the internal structure of  nucleon.
Based on this model, one can say that a nucleon consists of three valence quarks, surrounded by a large number of quark-antiquark pairs, called sea quarks $(u_s,\bar{u}_s,d_s,\bar{d}_s,\cdots)$ and  also gluons that makes the required interactions between quarks.
If we assume that the track of these pairs reach to the valence quarks, then one can say that three valence quarks are located in a sea of quarks and in the first approximation they have almost identical momentum distributions. Sea quarks are performed when gluons are considered as a double-colored identities. The inversion processes may be occurred and a pair quark anti-quark  annihilates to gluons. Therefore it can be said that we encounter with continuous stream of gluons. The sea quarks are in overall ``off-shell'' particles \cite{learning}.

{ Under the mentioned circumstances,} integration of  valence quark densities of the proton  over Bjorken-$x$ variable will lead to the following sum rules \cite{DIS}:
\begin{eqnarray}
&&\int_{0}^{1}\left[u(x)-\bar{u}(x)\right]dx=\int_{0}^{1}u_v(x)dx=2\;,\nonumber\\
&&\int_{0}^{1}\left[d(x)-\bar{d}(x)\right]dx=\int_{0}^{1}d_v(x)dx=1\;,\nonumber\\
&&\int_{0}^{1}\left[s(x)-\bar{s}(x)\right]dx=0\;.
\label{eq:1}
\end{eqnarray}
Above equations  are called valence sum rule which are corresponding to the two up valence and one down valence quark inside the proton.

Identification of partons in terms of quarks inside the proton with universal momentum distributions,  involves successful applications but this view has its own difficulties. By considering the experimental data and analysing  the theoretical estimations, one may conjecture that there should be other components in the proton which are carrying some part of  proton's momentum  but do not contribute in electromagnetic and  weak interactions. If all  momentum of proton are carried by  quarks and anti-quarks then one can write \cite{DIS}:
\begin{equation}
\sum_{i=1}^3\int_{0}^{1}dx\;x\;f_i(x)=1
\label{eq:3}\;.
\end{equation}
Experimental measurements do  not confirm the above relation  and the rest of the carried momentum  should be inevitably attributed to the gluons as another identity inside the proton. Considering gluons, the above relation will take the  following form:
\begin{equation}
\int_{0}^{1}dx\;x(\sum_{i=1}^3\;f_i(x)+g(x))=1\;.
\label{eq:4}
\end{equation}
If we assume the proton  at initial energy scale, $Q_0$,  is only composed of valence quarks as nonperturbative inputs, then Eq.(\ref{eq:4}) is represented by:
\begin{equation}
\int_{0}^{1}x[u_v(x,Q_0^2)+d_v(x,Q_0^2)]dx=1\;.
\label{eq:5}
\end{equation}
Correspondingly the quark number sum rule, given by Eq.(\ref{eq:1}), is written as:
\begin{eqnarray}
\int_{0}^{1}u_v(x,Q_0^2)dx&=&2\;,\nonumber\\
\int_{0}^{1}d_v(x,Q_0^2)dx&=&1\;.
\label{eq:2}
\end{eqnarray}
In coming sections, we take into account only Eq.(\ref{eq:5}) and Eq.(\ref{eq:2}) as the desired sum rules which help us, together with other constrains, to determine the unknown parameters of valence densities at initial energy scale. The other constrains back to the Heisenberg uncertainty  principle and maximum entropy principle  which we consider them in next two sections to employ in the parton model.

In fact the first part of this paper is essentially based on the method which was discussed in Ref.\cite{Chi-enropy}.
{ The present approach suffers from certain deficiencies, most notably the absence of the gluon density, which in turn inflates the number of free parameters in the parton density functions. To rectify this, we propose an improved methodology that exploits the full statistical information of the densities, as presented in Secs.~\ref{sec:7}, \ref{sec:8}, and \ref{sec:9}.}

\section{Heisenberg uncertainty principle and derivation of quark densities}\label{sec:3}

As we mentioned before, most part of this section and the next three ones are basically on the method which has been demonstrated in Ref.\cite{Chi-enropy}. However we improve this method later on in this paper but let us start exploitation this method as a first approach to determine the parton densities. In this regard we should say that according to the Heisenberg uncertainty principle, it is impossible that both the location and momentum of a particle can be determined precisely \cite{uncer}. The response of quantum theory is that such occasion  can be happened but not more accurate than the value that Heisenberg uncertainty principle is permitted, so:
\begin{equation}
\sigma_X \sigma_P\geq \hbar/2\;.
\label{eq:6}
\end{equation}
The above relation is applicable to a parton inside the nucleon. It is because that to each quark inside a proton, a probability density is assigned.
Here $\sigma_X$ is the standard deviation of { parton spatial position} in X direction and $\sigma_P$ is the standard deviation of a momentum accordingly. The first deviation, $\sigma_X$, is dependent on the radius of the proton.

{{ To be accurate, we ought to state that} the proton undergoes breakup in high-energy collisions, and a static geometrical radius is not well-defined. Indeed, in our approach, the uncertainty principle should not be used in a geometric sense (i.e., as a spatial size of a rigid sphere). Instead, it is interpreted statistically within a thermodynamic picture: the momentum spread of partons is related to an effective correlation length in the partonic medium at the initial resolution scale. In fact this correlation length should be understood as a decoherence or thermal wavelength scale, not the spatial radius of proton. Nevertheless we utilize effectively the  geometrical interpretation of proton to simplify the calculations at initial energy scale where the proton is in an static state.}

{ Following the strategy that is used in \cite{Chi-enropy}}, a simple estimation for $\sigma_X$  is obtained by  converting the spherical proton into a cylindrical one, without any change in the moving direction and volume of the proton.
{ In statistical mechanics, calculating parton distribution functions (PDFs) in spherical coordinates can lead to cumbersome integrals. By assuming a cylindrical geometry, the direction of the proton's momentum (the $X$-axis) becomes the axis of symmetry. This significantly simplifies the phase space calculations in momentum space. Bjorken-$x$ represents the longitudinal momentum fraction of the proton carried by a parton. The cylindrical geometry facilitates a direct mapping:
The longitudinal momentum can be associated with the length ($L$) of the cylinder.}

{Since the {proton} is supposed to move in X direction, the space component which is  effective for considering the  position uncertainty would be the longitudinal component of assumed  cylindrical shape of the { proton} that is aligned in X direction.}
Hence, to get the uncertainty of parton position in this direction, one can write \cite{Chi-enropy}:
\begin{equation}
\frac{4}{3}\pi R^3=\pi R^2 L\rightarrow\sigma_X=\frac{L}{2}=\frac{2}{3}R\;.
\label{eq:7}
\end{equation}
This is one way among the numerous methods to calculate $\sigma_X$. In this equation, $R=\sqrt{\langle r_p^2\rangle}$ is the average charge radius of the proton that has been measured in Lamb-Shift experiments and it is  about 0.841 fm \cite{po,an}.

{ To extend the above computation for the proton at the parton level for the $u$ quark type, we should assume the presence of two $u$ quarks that contribute equally to the spherical space of the proton, with each quark occupying half of the  total volume of proton. In this context, we should also recognize that $u$ valence quarks carry a positive electric charge, making it challenging for them to come very close to one another, which supports the assumption regarding the occupied volume of the proton by the two $u$ quark. Consequently, the radius of the spherical parton would be $\frac{R}{2^{1/3}}$ where $R$ denotes the radius of the spherical proton, taking into account the subsequent relationships:\\

$V_{Parton} =\frac{V_{Proton}}{2} $, $V_{Parton} =\frac{4}{3}\pi R_{Parton}^3$,
$\rightarrow$ $V_{Proton}=\frac{4}{3}\pi R^3=2*\frac{4}{3}\pi R_{Parton}^3$\\

$\Rightarrow$  $R_{Parton}=\frac{R}{2^{1/3}}$ \\

As mentioned earlier, to convert the spherical volume to a cylindrical volume for the proton moving in the $X$ direction, if we once more assume a cylindrical volume equivalent to the spherical shape for the $u$ quark, the resulting uncertainty for this quark will be as shown in Eq.(\ref{eq:7}), where $R$ is substituted with $\frac{R}{2^{1/3}}$. Thus, we obtain $\sigma_{X_u}=\frac{2 R}{3}\frac{1}{2^{1/3}}$ for the position uncertainty of the $u$ quark, while for the $d$ quark, which can fully occupy the entire volume of the proton, the uncertainty would be $\sigma_{X_d}=\frac{2 R}{3}$.\\

The results above can be associated with $x$ Bjorken variable, which, according to the quark-parton model, represents the momentum fraction of the proton that quark carries. Taking into account $\sigma_{X_u}$ and $\sigma_{X_d}$, as established earlier in spatial coordinate, then based on the Heisenberg uncertainty principle, $\sigma_P$ can be calculated in momentum space. Consequently, $\sigma_x=\frac{\sigma_P}{M_P}$ represents the uncertainty in x-Bjorken space, a dimensionless quantity evaluated at $Q_0$ as the initial energy scale, which is accessible. Here, the proton mass $M_P$ is approximately $0.938\; GeV$, and $\sigma_P$ is denoted as $\sigma_{P_u}$ and $\sigma_{P_d}$ for the up and down quarks, respectively, leading to $\sigma_{x_u}$ and $\sigma_{x_d}$ as the uncertainties related to the $x$-Bjorken variable.{}{ Here Heisenberg uncertainty relation is used only to estimate an effective volume in phase space for the partonic system in the rest frame of the nucleon. It does not imply that we are modelling the proton as a rigid shape in high-energy scattering.}}\\

On the other hand $\sigma_{x_u}$  and $\sigma_{x_d}$ can be computed in terms of expectation values of $x_{u}$ and  $x_{u}^{2}$ as are following:

\begin{eqnarray}
\sqrt{\langle x_{u}^{2}\rangle -\langle x_{u}\rangle ^{2}} &=&\sigma _{x_{u}}\;, \nonumber \\
\sqrt{\langle x_{d}^{2}\rangle -\langle x_{d}\rangle ^{2}} &=&\sigma _{x_{d}}\;.\label{expect}
\end{eqnarray}
The required expectation values in Eq.~(\ref{expect}) can be obtained in terms of quark densities, presenting below:
\begin{eqnarray}
\langle x_{u}\rangle  &=&\int_{0}^{1}x\frac{u_{v}(x,Q^{2})}{2}dx\;,  \nonumber
\\
\langle x_{u}^{2}\rangle  &=&\int_{0}^{1}x^{2}\frac{u_{v}(x,Q^{2})}{2}dx\;,
\nonumber \\
\langle x_{d}\rangle  &=&\int_{0}^{1}xd_{v}(x,Q^{2})dx\;,  \nonumber \\
\langle x_{d}^{2}\rangle  &=&\int_{0}^{1}x^{2}d_{v}(x,Q^{2})dx\;.  \label{eq:9}
\end{eqnarray}
{{}
Now, using Eq.(\ref{eq:9}) and considering  Eq.(\ref{expect}), we arrive at two relations which contain unknown parameters of valence densities.
In summary what will conduct us to the two relations in Eq.(\ref{expect}), is based on using the Heisenberg uncertainty principle. How to use these relations practically in our computations to extract parton densities, would be done in Sec.\ref{sec:5}}.

Before that we investigate the maximum entropy principle in next section  as another constrain which help us also to determine the unknown parameters of parton densities.
\section{Maximum entropy principle and non-perturbative input}\label{sec:4}

In order to use the  maximum entropy principle in quark-parton model, we first give a review on this principle.

The concept of entropy has been introduced firstly in theoretical physics by Clausius in the mid-19$^{th}$ century \cite{Cla}. The entropy of a system is a function of their thermodynamic coordinates and it is given by:
\begin{equation}
S_f-S_i=\int_{i}^{f}\frac{dQ}{T}\;.
\label{eq:10}
\end{equation}

The limits of above integral are placed between the initial and final state of system in a reversible path.
In Eq. (\ref{eq:10}) $T$ is denoting the temperature of system in equilibrium state and $dQ$ is the exchanged heat in system. Whenever a system, due to friction and some other reasons, is wasting its kinetic energy or losing the work which can do then the irregular motions of molecules are increasing. This kind of occurrence is indicating a transition from ordering to disordering. In classical thermodynamics, this means that entropy is increasing. Disordering  and entropy are changing parallel to each other. The irregularity of a system can be calculated, considering  the probability  that a system can reach to its macroscopic state. This thermodynamic probability can also be related to the disordering of system.

A macroscopic system has a small number of parameters to measure. They can be measured by ordinary tools but the number of parameters in the microscopic system are too many. So people are using the statistical mechanics  in macroscopic system to check the probability behaviour of a system. The statistical mechanics can be considered from   classical or quantum point of views. In both approaches, it is assumed that the particles of a system are indistinguishable and behaviour identically however we should note that identical particles in terms of the classical and the quantum mechanic are different \cite{Bin}. If we have quantum statics, the particles should obey from Fermi-Dirac or Bose-Einstein distributions, depending on their spin status. Just reminding that particles with integer spin is called boson and the ones with half-integer, named fermions. Considering the probability distributions of a thermodynamic  system, the entropy of a system can be defined by \cite{Weh}:
\begin{equation}
S=-k\sum_r {\left[ {{P_r}\ln {P_r} + (1 - {P_r})\ln (1 - {P_r})} \right]}\;.
\label{eq:11}
\end{equation}
{This equation includes additionally the last term compared to the related equation in \cite{Chi-enropy} to align with the Fermi-Dirac distribution.}
By increasing the number of available states, the probability function $P_r$, for each ${`\;}r^{\;,}$ state, tends to non-zero values and consequently the entropy of system is growing up. By increasing the number of states, most of the $P_r$  are getting small values. So, entropy increases significantly. For this reason, increasing the entropy and disordering of a system occur simultaneously and we lead to the principle of maximum entropy \cite{Weh}.

After a brief review of  maximum entropy principle, we are  now at the stage to employ  it in quark-parton model and to  extract the  valence quark distributions. This is one of the ways to take into account a certain type of information which is called ``Random Information'' \cite{Weh}. In this analysis, the known features of proton are random information which should be determined. The maximum entropy principle is the standard criterion to choose the best probability distribution function. The distribution will be the best that maximize the entropy function.

To employ the principle of maximum entropy in quark-parton model and optimize the  concerned entropy function, we differentiate  it with respect to the unknown parameter of parton densities. For this purpose we need  first to parameterize parton distributions  at initial energy scale, $Q_0$. Since QCD theory can not declare how to depend the distribution function on $x$, as Bjorken variable, therefore a parameterized function at the initial energy scale should be assigned them. The simplest primary form is given by $f(x,Q^2)=Ax^B(1-x)^C$. According to this parameterized form , the naive non-perturbative input for valence quarks are written as they follow:
\begin{eqnarray}
xu_v(x,Q^2)&=&{a_u}x^{b_u}(1-x)^{c_u}\;,\nonumber\\
xd_v(x,Q^2)&=&{a_d}x^{b_d}(1-x)^{c_d}\;.
\label{eq:12}
\end{eqnarray}

In next section we show how one can obtain the unknown parameters of above densities, using the Heisenberg uncertainty and also the maximum  entropy principle together with quark density sum rule and  also momentum sum rule.

\section{Extracting Unknown parameters, first approach}\label{sec:5}

We now at the position to extract the unknown parameters of the parton densities in Eqs.(\ref{eq:12}). For this purpose we consider the Heisenberg uncertainty principle, given by Eq.~(\ref{eq:6}) in minimum case such that:
\begin{equation}
\sigma _{X}\sigma _{P}=\hbar/2
\label{eq:13}
\end{equation}

In this equation $\sigma _{X}$ is denoting to spatial uncertainty and it is given by Eq.~(\ref{eq:8}) for  $d$ and $u$ quarks. By accessing to numerical value for $R$  as the radius of proton, the numerical amounts of  $\sigma _{X}$ for $d$ and $u$ quark can be obtained.  Now according to uncertainty principle, Eq.~(\ref{eq:13}), $\sigma _{P_{u}}$ and $\sigma _{P_{d}}$ as the momentum uncertainty of $d$ and $u$ quarks can be calculated numerically and following that using $\sigma_{x}=\frac{\sigma_{P}}{M_P}$, the uncertainty with respect to $x$-Bjorken variable, can also be computed numerically as it follows:

\begin{eqnarray}
\left.
\begin{array}{c}
\sigma _{x_{u}}=\frac{\sigma _{P_{u}}}{M_{P}} \\
M_{P}=0.938\;MeV%
\end{array}%
\right\} &\Longrightarrow &\sigma _{x_{u}}=0.23\;, \\
\left.
\begin{array}{c}
\sigma _{x_{d}}=\frac{\sigma _{P_{d}}}{M_{P}} \\
M_{P}=0.938\;MeV%
\end{array}%
\right\} &\Longrightarrow &\sigma _{x_{d}}=0.19\;.  \nonumber
\label{eq:14}
\end{eqnarray}
Taking into account Eq.~(\ref{expect}), the following numerical results will be obtained:
\begin{eqnarray}
\label{eq:15}
\langle x_{u}^{2}\rangle -\langle x_{u}\rangle ^{2} &=&0.0529\;, \\
\langle x_{d}^{2}\rangle -\langle x_{d}\rangle ^{2} &=&0.0361\;.  \nonumber
\end{eqnarray}

So far, according to Eqs.~(\ref{eq:5}, \ref{eq:2}, \ref{eq:15}) we have five equations to find the unknown parameters of non-perturbative functions in Eq.~(\ref{eq:12}). But we should be aware that each of these distributions has three unknown parameters. So we have totally 6 unknown parameters and five equations. Therefore, we need to another relation and for this case the concerned one is coming out from the principle of maximum entropy. According to Eq.~(\ref{eq:11}) and considering it for a continues distribution with respect to Bjorken $x$-variable the entropy function in terms of  valence densities, is as  following:

\begin{eqnarray}
\label{eq:16}
S&=&-k\int_{0}^{1}\bigg[ 2\frac{xu_{v}(x,Q_{0}^{2})}{2}\ln \left( \frac{xu_{v}(x,Q_{0}^{2})}{2}\right)+ 2\left(1-\frac{xu_{v}(x,Q_{0}^{2})}{2}\right)\ln \left( 1-\frac{xu_{v}(x,Q_{0}^{2})}{2}\right)\nonumber\\
&&+xd_{v}(x,Q_{0}^{2})\ln \bigg(xd_{v}(x,Q_{0}^{2})\bigg)+\bigg(1-xd_{v}(x,Q_{0}^{2})\bigg)\ln \bigg(1-(xd_{v}(x,Q_{0}^{2})\bigg) \bigg] dx\;.
\end{eqnarray}

Indicating the remaining unknown parameter in valence densities typically by $h$, the principle of maximum entropy leads to:
\begin{equation}
\label{eq:17}
\frac{\partial }{\partial h}S=0\;.
\end{equation}
Now, we have six equations and six unknown parameters in parton densities which are obtained by solving Eqs.~(\ref{eq:5}, \ref{eq:2}, \ref{eq:15}, \ref{eq:17}) simultaneously.
%For this purpose, we  use the  Mellin transformation on parton distribution to obtain their moments in Mellin $n$-space. Then, obtained all the equations in Melline space.
Numerical results for the distribution functions at initial energy scale ,$Q_0$, would be as they follow:
\begin{eqnarray}
xu_v(x,Q_0^2)&=&5.763x^{0.204}(1-x)^{1.147}\;,\nonumber\\
xd_v(x,Q_0^2)&=&5.528x^{0.295}(1-x)^{2.036}\;.
\label{eq:18}
\end{eqnarray}
These results are obviously satisfying the required sum rules.

By comparing  and matching the results for valence densities, given by Eq.~(\ref{eq:18}), with results from some phenomenological and parametrization models, the numerical value of initial scale, $Q_0$  would be obtained. Comparison with CT10 model~\cite{Lai:2010vv} lead us to $Q_0^2=0.4\; GeV^2$.

{In the applied approach, we defined the statistical entropy in the space of momentum states (phase space) and maximise it subject to physical constraints such as momentum conservation, parton number conservation, and normalisation. This maximisation yields a non-uniform distribution in momentum space. It should be noted that spatial uniformity (uniform density inside the nucleon) is a separate assumption and is distinct from uniformity in momentum space.
While it is true that maximum entropy is associated with a uniform distribution in coordinate space , in phase space, maximising the entropy under physical constraints naturally yields non-uniform distributions. This approach operates in momentum space, and the non-uniform distribution we obtain is not only appropriate but also consistent with the fundamental principles of statistical mechanics. Taking this reality into account, it is feasible to extend the findings for valence distributions at the initial energy scale to other higher scales, which will be addressed in the following section.}

\section{Evolution the valence densities and nucleon structure function \label{sec:6}}

Considering the  distributions in Eq.~(\ref{eq:18}) as the input densities and using Dokshitzer-Gribov-Lipatov-Altarelli-Parisi (DGLAP) evolution equations \cite{Dokshitzer:1977sg,Gribov:1972ri,Altarelli:1977zs} , the valence quark densities at high energy scale can be determined.

The evolution of valence densities is obtained basically according to \cite{Altarelli:1977zs}:
\begin{eqnarray}
\frac{\partial\; q(x,Q^2)}{\partial \log (Q^2)}=[p\otimes q](x,Q^2)\;.\label{evol}
\end{eqnarray}
Using Eq.~(\ref{evol}) to  evolve the parton density to energy scale $Q^2=10\; GeV^2$, the following results for parton densities at Leading order (LO)  accuracy are obtained:
\begin{eqnarray}
xu_v(x,Q_0^2)&=&3.448x^{0.702}(1-x)^{2.173}\;,\nonumber\\
xd_v(x,Q_0^2)&=&2.318x^{0.725}(1-x)^{3.382}\;.
\label{eq:19}
\end{eqnarray}
If we do the required calculations at the Next to leading order (NLO) accuracy, the numerical results for parton density at energy scale $Q^2=10\; GeV^2$ would  be as following:
\begin{eqnarray}
xu_v(x,Q_0^2)&=&2.303x^{0.55}(1-x)^{2.351}\;,\nonumber\\
xd_v(x,Q_0^2)&=&1.434x^{0.565}(1-x)^{3.544}\;.
\label{eq:20}
\end{eqnarray}

\begin{figure}[!htb]
	%	\vspace{-3.5cm}
	\begin{center}
	\includegraphics[clip,width=0.65\textwidth]{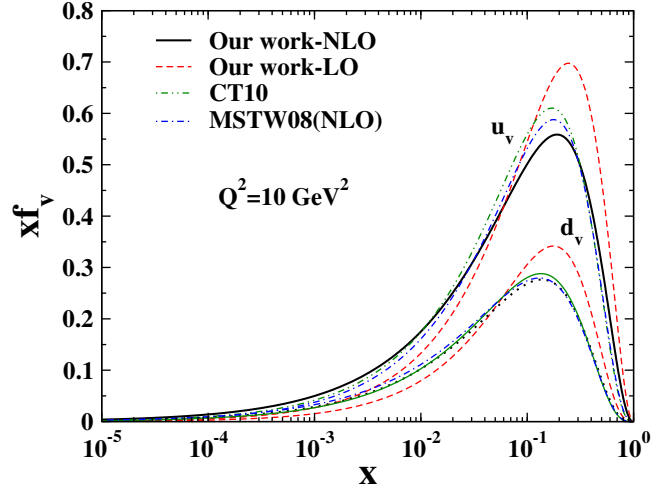}
	%\vspace{-6cm}
	%\label{u-d}
		\caption{{\small Comparison of $u$  and $d$ valence  densities at LO and NLO accuracies  with  CT10~\cite{Lai:2010vv} and MSTW08 ~\cite{Martin:2009iq} parametrization models, based on Heisenberg uncertainty  and maximum entropy principles. \label{fig:fig1}}}.
	\end{center}
\end{figure}
We depict the results of Eqs.~(\ref{eq:19},\ref{eq:20}) in Fig.\ref{fig:fig1} and compare them with result of CT10~\cite{Lai:2010vv} and MSTW08 ~\cite{Martin:2009iq} models. As it is seen our NLO results are in better agreement with the results from the parametrization models.

\begin{figure}[!htb]
	%	\vspace{-3.5cm}
	\begin{center}
	\includegraphics[clip,width=0.65\textwidth]{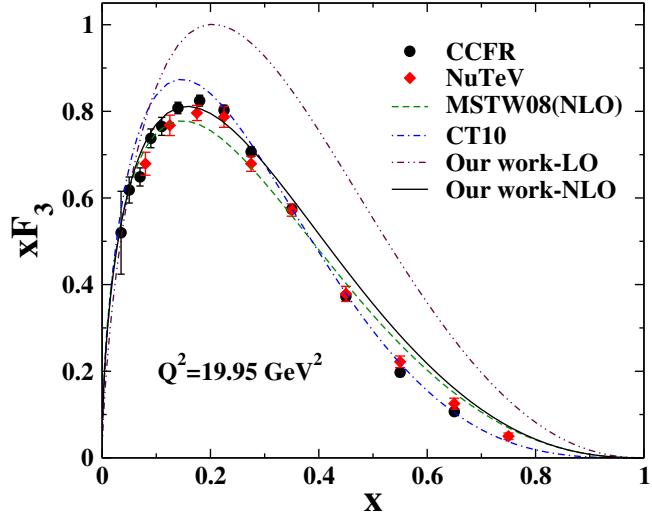}
	%\vspace{-6cm}
		\caption{{\small Structure function, $xF_3$, based on the Heisenberg  uncertainty and maximum entropy principles at the LO and NLO accuracies and its comparison with the available experimental  data \cite{Seligman:1997mc, NuTeV:2005wsg} and some parametrization models \cite{Martin:2009iq,Lai:2010vv}. \label{fig:fig2-5}}}
	\end{center}
\end{figure}

Now by accessing to the evolved parton densities we are at a position to calculate nucleon structure function. In this regard we should remind that
deep inelastic scattering of neutrino and antineutrino off the nucleon can be happened by weak interaction that is determined by two structure functions  where one of them is denoted by $xF_3$ \cite{DIS}. This nucleon structure function can be performed at the LO accuracy by taking the average of neutrino and antineutrino's interactions in terms of valence quark distributions as it follows \cite{DIS}:

\begin{eqnarray}
xF_3&=&\sum_q x (q(x,Q^2)-\bar{q}(x,Q^2))\nonumber\\
&=& x (u_v(x)+d_v(x))
\label{eq:21}\;.
\end{eqnarray}
{ Considering the evolved valence distributions at a different energy scale, specifically $Q^2=20\; GeV^2$, which allows for a comparison with the availabe experimental data, the following result for $xF_3$ is derived:}
\begin{eqnarray}
xF_3&=&3.285x^{0.673}(1-x)^{2.257}\nonumber\\
&+&2.169x^{0.692}(1-x)^{3.473}\;.
\label{eq:22}
\end{eqnarray}
Considering  Eq.~(\ref{eq:22}) which is based on employing the Heisenberg uncertainty and Maximum entropy principles, we are able to plot $xF_3$ structure function at the LO accuracy which is depicted in Fig.\ref{fig:fig2-5}.

To evaluate the $xF_3$ structure  function at the NLO accuracy, we  utilize the  following relation \cite{DIS}:

\begin{eqnarray}
xF_3&=&\sum_q x (q(x,Q^2)-\bar{q}(x,Q^2))\nonumber\\
&&+\frac{\alpha_s(Q^2)}{4 \pi}[C_i^1\otimes (q(x,Q^2)-\bar{q}(x,Q^2))]\;.
\label{eq:266}
\end{eqnarray}
In this equation $C_i^1$ is denoting the Wilson coefficient function at the NLO accuracy.  The required information about the coefficient function can be found in \cite{thesis-35}.
Doing the convolution integral in Eq.~(\ref{eq:266}) we could compute the $xF_3$ structure-function numerically in Mellin  $n$-space and finally the results  in Bjroken $x$-space would be obtained. We plot the result of  Eq.(\ref{eq:266}) for $xF_3$ in  Fig.\ref{fig:fig2-5} wherein the results for  structure function at LO and  NLO accuracies are compared with CCFR ~\cite{Seligman:1997mc} and NuTev experimental data ~\cite{NuTeV:2005wsg} and also with MSTW ~\cite{Martin:2009iq} and CT10~\cite{Lai:2010vv} parametrization models  which are indicating that the NLO results are in good  agreement with them.\\

\begin{figure}[!htb]
	%	\vspace{-3.5cm}
	\begin{center}
	\includegraphics[clip,width=0.65\textwidth]{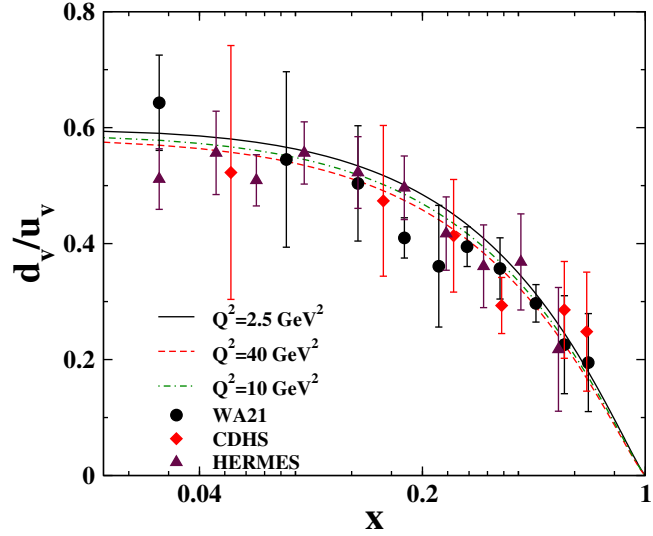}
	%\vspace{-6cm}
		\caption{{\small The ratio of $d$ and $u$ valence densities at three different energy scales  and their comparison with the available experimental data \cite{Birmingham-CERN-ImperialCollege-MunichMPI-Oxford:1994pnp,Abramowicz:1984yk,HERMES:1997}, based on the Heisenberg uncertainty and maximum entropy principles.\label{fig:fig2}}}
	\end{center}
\end{figure}

{ Another measure utilized to signify the effectiveness of the employed method in determining the valence quark distributions is the ratio of valence densities, $\frac{d_v}{v_v}$.} We calculate it at three energy scales $Q^2=2.5, 10$ and $40\;GeV^2$
The numerical expressions for the ratio at mentioned energy scales are as following respectively:
\begin{eqnarray}
R_{2.5}&=&0.6457x^{0.0154279}(1-x)^{1.176652}\;,\nonumber\\
R_{10}&=&0.622117x^{0.0105865}(1-x)^{1.193135}\;,\nonumber\\
R_{40}&=&0.60704x^{0.0076785}(1-x)^{1.207516}\;.
\label{eq:24}
\end{eqnarray}
We depict these ratio with respect to $x$ variable in Fig.\ref{fig:fig2} and compare them with the WA21 ~\cite{Birmingham-CERN-ImperialCollege-MunichMPI-Oxford:1994pnp}, CDHS ~\cite{Abramowicz:1984yk} and HERMES ~\cite{HERMES:1997} experimental data. { As it is expected, these ratios exhibit minimal dependence on energy scales and align well with existing experimental data, which further demonstrates the effectiveness of Heisenberg uncertainty and maximum entropy principles in determining the valence quark densities at the initial energy scale.}

\section{Quark model in statistical (second) approach\label{sec:7}}
{ So far, we have successfully derived the valence quark densities using Heisenberg uncertainty and maximum entropy principles, which, as previously illustrated, builds on the work done in Ref.\cite{Chi-enropy} with certain modifications; we refer to this as the first approach.
In this approach, the gluon density cannot be determined, and as we increase the number of parton densities, the count of unknown parameters rises, leading to a scenario where this approach fails to yield the required parton densities. Given these limitations, we introduce a different method, referring to it as the second approach, which allows us to extract the gluon density while while maintaining the number of unknown parameters at four.}

In this (second) approach we intend to extract the quark and gluon densities based on the statistical method which consists some unknown parameters like the volume, $V$,  temperature, $T$, and chemical potentials, $\mu$, of considered system for nucleon. We try to obtain these parameters using the quark number and momentum  sum rules together only and just with the principle of maximum entropy. Here we do not resort to the principle of Heisenberg uncertainty since there is not necessary to use it in considering the statistical method. Before to utilize this approach we give a brief discerption for it as in following.

Since quarks and gluons are confined inside the  nucleon, it is expected that statistical properties are important in determining their distribution functions. In this approach, a proton is assumed to be a system in thermal equilibrium which consists of free partons.

{}{In fact the proton in a high-energy DIS event is not literally an equilibrium thermodynamical system, nor an ideal gas, in the dynamical sense of the interaction. However, it should be clarified  that the assumption of an ideal gas at equilibrium is not intended as a physical reality of the collision process itself, but rather as a statistical prior or maximum-entropy initial condition for the parton distribution at a low reference scale. Indeed, the ideal-gas equilibrium model is used here as a minimum-bias statistical prior for the low-scale PDF, not as a claim about the physical state of the proton during or after the hard collision.}

The history of this research area  backs to 1990 when Qi-bo Ma and collaborations  considered the nucleon as a statistical system, composed of free partons \cite{thesis-59,thesis-58,sof1,sof2}. For this purpose they assigned to partons a Fermi-Dirac statistical distribution and by using a desired transformation they could do the required calculations in the light cone coordinates. Later on the parton statistical model has been investigated with more details in \cite{ma-EMC} and its applications was extended by some people \cite{MDY}.

Employing the quark-parton model in QCD theory has been encountered primarily with some difficulties. It is because that the perturbative section of theory  was well tested  and considered while its non-perturbative section was not noticed and investigated  well like the perturbaive one. The best way to establish a link between the two sections was provided by the light front dynamics. In fact, the main formulation was achieved in these coordinates and this idea led to a quark model that has a light front structure. In this connection let us first introduce the kinematic coordinates in proton frame, using the light cone coordinates \cite{thesis-58}:
\begin{eqnarray}
\label{eq:25}
k_{i}^{\mu } &=&(k_{i}^{+},k_{i}^{-},k_{\perp i})\;, \\
k_{i}^{+} &=&x_{i}P^{+} \label{kk}\;, \\
p_{\perp i} &=&x_{i}P_{\perp}+k_{\perp i}\;,
\end{eqnarray}
where  $P^{\pm}=P^0\pm P^3$ and $x_i$ is the longitudinal momentum fraction, carried by  $i^{th}$ quark. Here $P^{+}$  consists a sum of proportional component of $n$ quarks such as:
\begin{equation}
\label{eq:26}
P^{+}=\sum_{i}^{n}k_{i}^{+}\;\;\;,\;\;\;k_{i}^{+}>0\;.
\end{equation}
Considering Eq.~(\ref{kk}) and (\ref{eq:26}), as it is expected, we arrive at the following result:
\begin{eqnarray}
\label{eq:27}
\sum_{i}^{n}x_{i} &=&1\;.
\end{eqnarray}
Since $P_{\perp}=\sum_{i}^{n} p_{\perp i}$ and considering Eq.~(\ref{eq:27}) we conclude that
\begin{eqnarray}
\sum_{i}^{n}k_{\perp i} &=&0\;.
\end{eqnarray}
Considering the mass shell condition $k_i^2=m_i^2$ and also the scalar product of $k_i$ four vector, $k_{\mu i}k_{i}^{\mu }=k_{i}^{+}k_{i}^{-}-k_{\perp i}^{2}$, each single particle state would have  the following four momentum:
\begin{equation}
\label{eq:28}
k_{i}^{\mu }=(k_{i}^{+},k_{i}^{-},k_{\perp i})=(k_{i}^{+},\frac{k_{\perp
		i}^{2}+m_{i}^{2}}{k_{i}^{+}},k_{\perp i}),\;\;i=1,2,\cdots ,n
\end{equation}
Then the invariant mass of the constituents in each n-particle state is given by:
\begin{eqnarray}
\label{eq:29}
M_{n}^{2} &=&\left( \sum_{i=1}^{n}k_{i}^{\mu }\right) ^{2}=\left(
\sum_{i=1}^{n}k_{i}^{+}\right) \left( \sum_{i=1}^{n}k_{i}^{-}\right) -\left(
\sum_{i=1}^{n}k_{\perp i}\right) ^{2}\nonumber \\
&=&\sum_{i=1}^{n}\frac{m_{i}^{2}+k_{\perp i}^{2}}{x_{i}}\;.
\end{eqnarray}

Since we are familiar with the  basic tools to consider light cone frame  in our calculations, we are now at the stage to utilize these coordinates in the statistical approach as it describes in coming section.

\section{Statistical distribution in light cone coordinates \label{sec:8}}

From statistical point of view, a nucleon is an ideal gas system consisting of  $n$-particle, composed of partons (quark, anti-quark and gluon) in a thermal equilibrium at $T$ temperature in a volume of $V$. Ideal gas is a system which its particles do not interact with each other.

Using the statistical considerations, average number of particles in a system is given by \cite{Bin}:
\begin{equation}
\label{eq:31}
\langle n\rangle =\frac{1}{e^{y}\pm 1}\;.
\end{equation}
Here positive sign is denoting to fermions and negative sign to bosons. The $y$ parameter is defined as it follows:	
\begin{equation}
\label{eq:32}
y\equiv \frac{\varepsilon -\mu }{kT}\;,
\end{equation}
where $\mu$ is the particle chemical potential and $\varepsilon$ is energy eigenvalue of system in a specified state.

Following this procedure for a continues system, the average number of particles considering the related phase space, would be represented by:
\begin{equation}
\label{eq:34}
\overline{N}_{f}=\int f(k^{0})d^{3}k\;.
\end{equation}
In this relation the energy-momentum four-vector, $k$, has the following components:
\begin{eqnarray}
\label{eq:35}
k^{0} &=&\sqrt{K^{2}+m_{f}^{2}}\;, \\
K &=&(k^{1},k^{2},k^{3})\;,  \nonumber
\end{eqnarray}
where $k^0$, $K$ and $m$ are energy, spatial momentum vector and particle mass respectively. The function, $f(k^0)$, is denoting distribution density in $V$ volume and is given by \cite{Weh}:
\begin{equation}
\label{eq:36}
f(k^{0})=\frac{V}{(2\pi )^{3}}\frac{1}{e^{\frac{k^{0}-\mu }{T}}\pm 1}
\end{equation}
Now, if we assume nucleon is a system in thermal equilibrium, $f(k^0)$ will be density function. In case that there is a degeneracy , the above distribution is given by:
\begin{equation}
\label{eq:37}
f(k^{0})=\frac{g_{f}V}{(2\pi )^{3}}\frac{1}{e^{\frac{k^{0}-\mu }{T}}\pm 1}\;,
\end{equation}
where $g_f$ is called degeneracy factor which is equal to 6 for quarks and anti-quarks as fermions and 16 for gluon that is boson . In above equation $\mu$ is denoted to chemical potential which has the following properties \cite{chem-1,chem-2,chem-3,chem-4} :
\begin{eqnarray}
\label{eq:38}
\mu _{\overline{q}} &=&-\mu _{q}\Longrightarrow \left\{
\begin{array}{c}
\mu _{\overline{u}}=-\mu _{u}\;, \\
\mu _{\overline{d}}=-\mu _{d}\;,%
\end{array}%
\right. \\
\mu _{g} &=&0\;.  \nonumber
\end{eqnarray}

According to the principle of energy-momentum conservation, Eq.~(\ref{eq:34}) will appear as it follows
\begin{eqnarray}
\label{eq:39}
\overline{N}_{f} =\int f(k^{0})\delta (k^{0}-\sqrt{(k^{3})^{2}-k_{\perp
	}^{2}+m_{f}^{2}})dk^{0}dk^{3}d^{2}k_{\perp} \;,\nonumber\\
\end{eqnarray}
where one can write
\begin{eqnarray}
\delta (k^{0}-\sqrt{(k^{3})^{2}-k_{\perp }^{2}+m_{f}^{2}}) =2k^{0}\theta
(k^{0})\delta (k^{2}-m_{f}^{2})\;.  \nonumber\\
\end{eqnarray}

Now, we will use the light cone coordinates and transform the variables into the kinetic variables in the light cone coordinate system using the following transformations:
\begin{eqnarray}
\label{eq:40}
k^{+} &=&k^{0}+k^{3}\;\;,\;\;k^{-}=k^{0}-k^{3}\;, \\
k_{\perp } &=&(k^{1},k^{2})\;\;,\;\;k^{+}=P^{+}x=Mx\;.  \nonumber
\end{eqnarray}
Here $x$ is the momentum fraction of nucleon,  carried by partons in light cone coordinates and $M$ is proton mass. Using light front variables, delta Dirac function and  measure of integration in Eq.~(\ref{eq:39}) will change respectively to \cite{thesis-59,thesis-58,MDY}:

\begin{eqnarray}
\label{eq:41}	
2k^{0}\theta (k^{0})\delta (k^{2}-m_{f}^{2}) &=&\left[ 1+\frac{%
	k_{\perp }^{2}+m_{f}^{2}}{Mx^{2}}\right] \\
&&\theta (k^{0})\delta \left( k^{-}-\frac{k_{\perp }^{2}+m_{f}^{2}}{Mx}%
\right)\;,   \nonumber
\end{eqnarray}

\begin{equation}
\label{eq:42}
dk^{0}dk^{3}d^{2}k_{\perp }=\frac{1}{2}Mdk^{-}dxd^{2}k_{\perp }\;.
\end{equation}
Substituting Eqs.~(\ref{eq:41}, \ref{eq:42}) into Eq.~(\ref{eq:39}) and doing integration with respect to $k^-$, one will arrive at:

	\begin{eqnarray}
	\label{eq:43}
	\overline{N}_{f} &=&\int f(x,k_{\perp })dxd^{2}k_{\perp }\;, \\
	f(x,k_{\perp }) &=&\frac{g_{f}MV}{2(2\pi )^{3}}\frac{1}{\exp \left( \frac{%
			\frac{1}{2}\left( Mx+\frac{k_{\perp }^{2}+m_{f}^{2}}{Mx}\right) -\mu _{f}}{T}%
		\right) \pm 1}\left[ 1+\frac{k_{\perp }^{2}+m_{f}^{2}}{Mx^{2}}\right] \theta
	(x)\;.  \nonumber
	\end{eqnarray}	
	By integrating the above equation with respect to $k_{\perp }$ which is assumed to be isotropic  in  transverse plane, the distribution function in terms of Bjorken variable $x$ is finally coming out as it follows:
	\begin{equation}
	\label{eq:44}
	f(x)=\pm \frac{g_{f}MV}{2(2\pi )^{3}}\left\{ \ln \left[ 1\pm \exp \left( -%
	\frac{\frac{1}{2}\left( Mx+\frac{m_{f}^{2}}{Mx}\right) -\mu _{f}}{T}\right) %
	\right] -2TLi_{2}\left( \mp \exp \left( -\frac{\frac{1}{2}\left( Mx+\frac{%
			m_{f}^{2}}{Mx}\right) -\mu _{f}}{T}\right) \right) \right\}\;.
	\end{equation}
	Here the positive sign is referring to fermions and  negative sign to bosons while we take into account the properties of chemical potential, presented by Eq.(\ref{eq:38}), in our computations.
	
	Now according to Eq.~(\ref{eq:44}) and ignoring the quark mass, the parton distributions including quarks as $u(x),\bar{u}(x),d(x),\bar{d}(x)$ and gluon  density as $g(x)$ would be obtained as in following:
	\begin{eqnarray}
	\label{eq:45}
	u(x) &=&\frac{3MV}{(2\pi )^{3}}\left\{ \ln \left[ 1+\exp \left( -\frac{\frac{%
			1}{2}\left( Mx\right) -\mu _{u}}{T}\right) \right] -2TLi_{2}\left( -\exp
	\left( -\frac{\frac{1}{2}\left( Mx\right) -\mu _{u}}{T}\right) \right)
	\right\}\;, \nonumber \\
	\overline{u}(x) &=&\frac{3MV}{(2\pi )^{3}}\left\{ \ln \left[ 1+\exp \left( -%
	\frac{\frac{1}{2}\left( Mx\right) +\mu _{u}}{T}\right) \right]
	-2TLi_{2}\left( -\exp \left( -\frac{\frac{1}{2}\left( Mx\right) +\mu _{u}}{T}%
	\right) \right) \right\}\;,   \nonumber \\
	d(x) &=&\frac{3MV}{(2\pi )^{3}}\left\{ \ln \left[ 1+\exp \left( -\frac{\frac{%
			1}{2}\left( Mx\right) -\mu _{d}}{T}\right) \right] -2TLi_{2}\left( -\exp
	\left( -\frac{\frac{1}{2}\left( Mx\right) -\mu _{d}}{T}\right) \right)
	\right\}\;,   \nonumber \\
	\overline{d}(x) &=&\frac{3MV}{(2\pi )^{3}}\left\{ \ln \left[ 1+\exp \left( -%
	\frac{\frac{1}{2}\left( Mx\right) +\mu _{d}}{T}\right) \right]
	-2TLi_{2}\left( -\exp \left( -\frac{\frac{1}{2}\left( Mx\right) +\mu _{d}}{T}%
	\right) \right) \right\}\;, \nonumber \\
	g(x) &=&-\frac{MV}{(\pi )^{3}}\left\{ \ln \left[ 1+\exp \left( -\frac{\frac{1%
		}{2}\left( Mx\right) }{T}\right) \right] -2TLi_{2}\left( -\exp \left( -\frac{%
		\frac{1}{2}\left( Mx\right) }{T}\right) \right) \right\}\;. \nonumber\\
	\end{eqnarray}

In these equations $Li_{2}$ is polylogarithm function. {There are four free parameters in the equations above: T, V, and the chemical potentials $\mu_u$, $\mu_d$, which will be determined using the quark number and momentum sum rules along with the principle of maximum entropy, as we will explain in the following section.}

\section{Sum rules and  principle of maximum entropy in statistical approach}\label{sec:9}

Unknown parameters that we consider in statistical distribution functions are $T$, $V$ and chemical potential $\mu_u$ and $\mu_d$. As pointed out before, to obtain these unknown parameters we need some  constraints, including quark number and momentum sum rules together with the maximum entropy principle. For this purpose, we should take into account these constraints and solve the equations which are arising them simultaneously.

The momentum sum rule in the presence of quark, anti-quark and gluon densities is  as it follows:
\begin{equation}
\label{eq:46}
\int_{0}^{1}x[u(x)+\overline{u}(x)+d(x)+\overline{d}(x)+g(x)]dx=1
\end{equation}	
The sum rules that represent the number of valence quarks are given by Eq.(\ref{eq:2}).

{The additional equation we require is the maximum entropy principle. In this instance, it is necessary to expand Eq.(\ref{eq:16}) to also encompass anti-quark and gluon densities. It is important to mention that using Eq.(\ref{eq:16}) or, similarly, Eq.(\ref{eq:11}) to account for gluon density must incorporate its second term under integration with a positive sign within the parentheses, as gluons, being bosons, follow the Bose-Einstein distribution, unlike valence and sea quarks that follow the Fermi-Dirac distribution.}

Since we have four unknown parameters, we consider optionally the $\mu_d$ as a free parameter in which the entropy, $S$, should be optimized with respect to it:
\begin{equation}
\label{eq:49}
\frac{\partial }{\partial \mu _{u}}S=0\;.
\end{equation}
Considering the above equation together with quark number and  momentum sum rules, Eqs.(\ref{eq:5},\ref{eq:2},\ref{eq:46}), and solving these equations simultaneously we will arrive at the following numerical results:    :
\begin{eqnarray}
\label{eq:50}
T &=&49.457\;\;,\;\;V=8.979\times 10^{-6}\;, \\
\mu _{d} &=&41.704\;\;,\;\;\mu _{u}=73.313  \nonumber
\end{eqnarray}
By replacing the above numerical values in  Eq.~(\ref{eq:45}), initial parton densities with respect to the Bjorken variable $x$ will be finalized as:

	\begin{eqnarray}
	\label{eq:51}
	u(x) &=&30.008x\ln [1+e^{(1.467-9.391x)}]-3.195Li_{2}[2,-e^{(1.467-9.391x)}]
	\\
	d(x) &=&30.008x\ln [1+e^{(0.835-9.391x)}]-3.195Li_{2}[2,-e^{(0.835-9.391x)}]
	\nonumber \\
	\overline{u}(x) &=&30.008x\ln
	[1+e^{(-1.467-9.391x)}]-3.195Li_{2}[2,-e^{(-1.467-9.391x)}]  \nonumber \\
	\overline{d}(x) &=&30.008x\ln
	[1+e^{(-0.835-9.391x)}]-3.195Li_{2}[2,-e^{(-0.835-9.391x)}]  \nonumber \\
	g(x) &=&80.02x\ln [1+e^{(-9.391x)}]-8.521Li_{2}[2,-e^{(-9.391x)}]  \nonumber
	\end{eqnarray}

We are now at the situation to use the obtained results for statistical parton densities and to do some phenomenological inspections. We deal with them in next section.

\section{Phenomenological achievements in statistical approach}\label{sec:10}
To obtain the distribution functions at the high energy scales, we need to evolve them. For doing this, we use  DGLAP evolution equations in two LO and NLO accuracies. At first and in the LO accuracy, we evolve the distribution functions to $Q^2=10\;GeV^2$ scale which can be done numerically. For this purpose we evolve the Mellin moments of parton densities at energy scale $Q_0^2$ to desired scale $Q^2$. The initial parton densities as the input of evolution processes is given by Eq.(\ref{eq:51}). In  this processes we take into account $\Lambda=0.248\;GeV$ as the cut off QCD parameter and the number of active quark flavour is considered to be $N_f=3$. As we mentioned before, by comparing the initial results for parton densities with CT10 parametrization model \cite{Lai:2010vv} we could estimate $Q^2_0=0.4\;GeV^2$  as the initial energy scale. We then make a data table from the evolved moments in terms of their orders, Afterwards we fit the  data table to the moment of  parton densities which involve the standard form  as  $f(x)=Ax^B(1-x)^C$ in Bjorken-$x$ space. The results of the fit  provide us the valence quark and gluon distributions at the evolved $Q^2$  scale. Their numerical presentations are as following:
\begin{eqnarray}
\label{eq:52}
xu_{v} &=&1.936x^{0.349}(1-x)^{3.737}\;, \\
xd_{v} &=&0.904x^{0.324}(1-x)^{3.787}\;,  \nonumber \\
xg &=&0.549x^{-0.439}(1-x)^{5.638}\,.  \nonumber
\end{eqnarray}

\begin{figure}[!htb]
	%	\vspace{-3.5cm}
	\begin{center}
	\includegraphics[clip,width=0.65\textwidth]{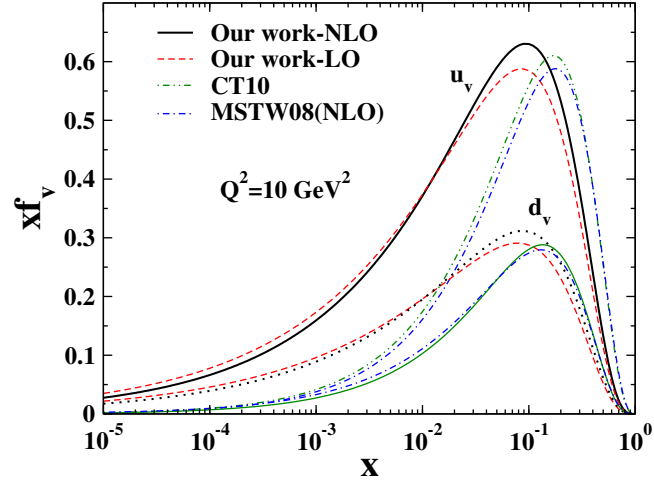}
	%\vspace{-6cm}
		\caption{{\small Quark valance densities in statistical approach which are compared with some parametrization models. \cite{Lai:2010vv,Martin:2009iq} \label{fig:fig3}}}
	\end{center}
\end{figure}

\begin{figure}[!htb]
	%	\vspace{-3.5cm}
		\begin{center}
	\includegraphics[clip,width=0.65\textwidth]{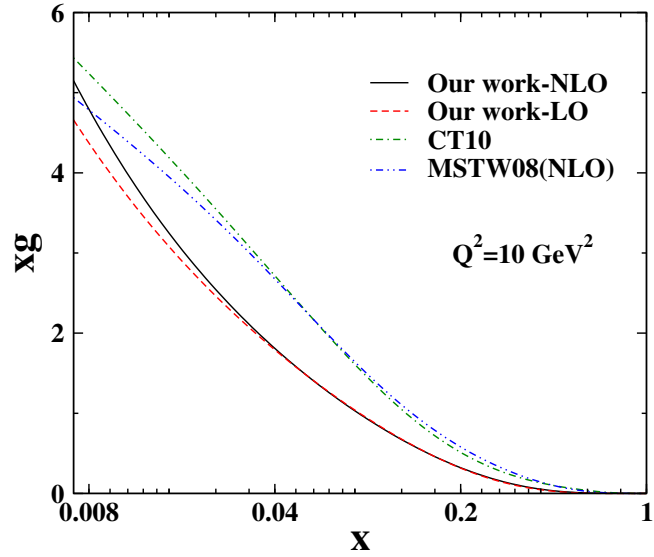}
	%\vspace{-6cm}
		\caption{{\small Gluon density in statistical approach and its comparison with some parametrization models \cite{Lai:2010vv,Martin:2009iq} \label{fig:fig4}}}
	\end{center}
\end{figure}

The same procedure can be done at the NLO accuracy in which the results in Eq.(\ref{eq:51}) are taken as  the input distributions at initial energy scale, $Q_0$, but in continuation the evolution processes in Mellin moment space is done at the NLO accuracy. The outcome of the evolution are the parton densities at the NLO accuracy in Bjorken-$x$ space with the following numerical results:
\begin{eqnarray}
\label{eq:58}
xu_{v} &=&2.238x^{0.382}(1-x)^{3.681}\;, \\
xd_{v} &=&1.044x^{0.356}(1-x)^{3.730}\;,  \nonumber \\
xg &=&0.442x^{-0.502}(1-x)^{5.064}\;.  \nonumber
\end{eqnarray}
The LO and NLO evolved parton densities in Eqs.(\ref{eq:52},\ref{eq:58}) are plotted in Figs.(\ref{fig:fig3},\ref{fig:fig4}) and compared with C10 and MSTW08 parametrization  models \cite{Lai:2010vv,Martin:2009iq} which indicate acceptable behaviour. As can be seen, the NLO results  are more compatible with parametrization models and at large $x$ values the plots tend to each other.
\begin{figure}[!htb]
	%	\vspace{-3.5cm}
		\begin{center}
	\includegraphics[clip,width=0.65\textwidth]{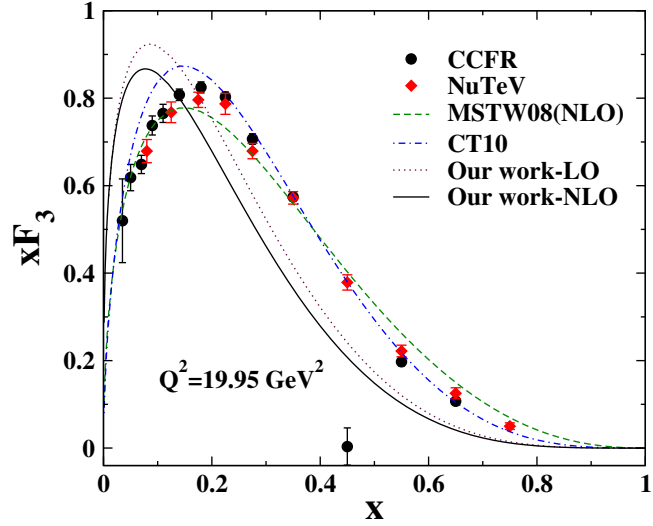}
	%\vspace{-6cm}
		\caption{{\small Structure function $xF_3$ based on  the statistical approach at LO and NLO accuracies and its comparison with some experimental data \cite{Seligman:1997mc, NuTeV:2005wsg} and parametrization CT10 \cite{Lai:2010vv} and MSTW08 \cite{Martin:2009iq} models. \label{fig:fig5}}}
	\end{center}
\end{figure}
\begin{figure}[!htb]
	%	\vspace{-3.5cm}
		\begin{center}
	\includegraphics[clip,width=0.45\textwidth]{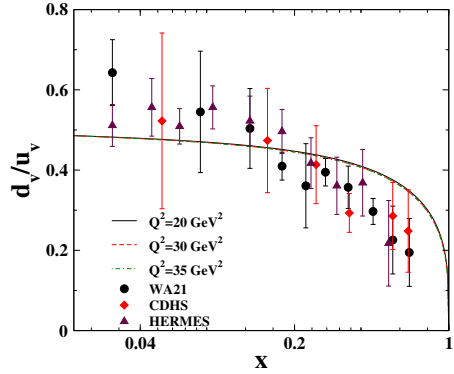}
	%\vspace{-6cm}
		\caption{{\small The ratio of the valance quark densities in statistical approch at three different energy scales and its comparison with WA21 ~\cite{Birmingham-CERN-ImperialCollege-MunichMPI-Oxford:1994pnp}, CDHS ~\cite{Abramowicz:1984yk} and HERMES ~\cite{HERMES:1997} experimental data. \label{fig:fig6}}}
	\end{center}
\end{figure}

Now, let us look at the $xF_3$ structure function. We could compute this function at $Q^2=20\; GeV^2$ in both LO and NLO accuracies (see Eqs.(\ref{eq:21},\ref{eq:266})). The final results are shown in Fig.\ref{fig:fig5}. Considering this figure, it is seen  that at small and large $x$ values, as we expected, this function goes to zero. We  also compare the determined functions with MSTW08 parametrization  models \cite{Lai:2010vv,Martin:2009iq}  and also with the available experimental data \cite{Seligman:1997mc,NuTeV:2005wsg}. The LO and NLO results of statistical approach have less difference with respect to the  results from first approach and on the other hand there is a little difference between the obtained results of the second (statistical) approach  and the available experimental data, as shown in Fig.\ref{fig:fig5}. One of the reasons for this  discrepancy, is that in current analysis we only resort to theoretical investigations and  do not use any experimental data in our calculations but in the first method, based on Heisenberg uncertainty and maximum entropy principles, we utilize the  radius of proton, measured from the Lamb-Shift process, as an experimental input.

As a final phenomenological achievement in parton statistical approach, we calculate the ratio of evolved valence quarks at $Q^2=20,30$ and $35\;GeV^2$  and plot them in Fig.\ref{fig:fig6}. We compare them with the CDHS ~\cite{Abramowicz:1984yk}, HERMES ~\cite{HERMES:1997} and WA21 ~\cite{Birmingham-CERN-ImperialCollege-MunichMPI-Oxford:1994pnp} experimental data. As can be seen, the ratio in $0.05<x<0.5$ interval has a better consistency  with the experimental data.
{ The outcome also shows an absence of reliance on the selected energy scale, as we anticipated.}
\section{Discussion and conclusion \label{sec:11}}
Extracting valence quark densities, using maximum entropy principle helps us to understand the primary features of the nucleon structure and to search for more details of the structure function. In this work the quark densities have been achieved by two different approaches. In the first approach, to compute the valence quark distribution functions, we allocated them a parametrized form at initial energy scale, $Q_0$, as non-perturbative input which included several parameters. These unknown parameters  have been computed, using the quark number and momentum sum rules together with principles of maximum entropy and  Heisenberg uncertainty. We then evolved the obtained parton densities to high energy scales and could calculate the ratio of valence densities at different energy scales and also the $xF_3$ nucleon structure function which indicate proper behaviour and are in good agreement with some parametrization models and also the available experimental data. The results of this approach have been depicted in Figs.\ref{fig:fig1},\ref{fig:fig2-5},\ref{fig:fig2}.

{ Current PDFs fits are usually based on employing available data to find the unknown parameters of parton densities. If we could determine all these parameters or part of them without to resort  experimental data and only utilize some constrains, arising out from principles such as Heisenberg uncertainty and maximum entropy together with several sum rules as momentum and quark number conservations, then the obtained results have more worthwhile from theoretical point of view. This procedure can be considered as an alternative method with respect to  underlying fitting method  to extract the PDFs.

On the other hand, when there is an intricate parametrization, it is required to incorporate, in addition to these principles, some additional constraints which might lead to better results for PDFs. Such constraint, for instance, could be the Gross-Llewellyn Smith sum rule \cite{GLS} , which provides valuable information on valence quark distributions at high $Q^2$. Another useful constraint is the Ellis-Jaffe sum rule \cite{EGS} , which can be applied to restrict polarized PDFs in practice.

Moreover these principles can be applied easily to other types of PDFs, including polarized PDFs, generalized parton distributions (GPDFs), and TMDPDFs as transverse momentum dependent parton densities. In such situations where comparatively little information is available from QCD calculations, the introduced method can prove particularly useful for obtaining reasonable results for PDFs that can be considered  as an extension to realize these principles.}

In the second approach we considered a statistical approach in which partons, including quarks and gluon, are assumed  as free particles which perform a thermal equilibrium system. Considering the constraints arising out from the maximum entropy principle and the current sum rules without to resort to the Heisenberg uncertainty principle, the initial distribution functions  in terms of statistical parameters as T, V and chemical potentials have been determined. Like the first approach we evolved the initial parton densities to high energy scales and  in addition to valence quark and gluon densities  at these scales we computed also the ratio of valence densities together with $xF_3$ nucleon structure functions. The results of this approach have been depicted in Figs.\ref{fig:fig3},\ref{fig:fig4},\ref{fig:fig5},\ref{fig:fig6}. Since in this method the evolved gluon distribution has also been obtained, we expected to achieve to more reliable results with respect to first approach.

The second approach gets priority with respect to first one. The first reason is that while we use in the first approach the principle of maximum entropy which is essentially a thermodynamic tool but we  use just from this thermodynamic feature and do not use  the other statistical features of parton densities. So we could claim that in second approach we are placed at a proper position to investigate the full statistical  features of parton system. Secondly in the second approach there is an opportunity to extract the gluon density function in our computations which is vital for our calculations at the NLO accuracy while there is not such situation in the first approach. Thirdly in the second approach in spite that we achieve to more outputs but we should determine less unknown parameters which can be considered as an advantage of this approach. So we expect  to get in second approach more reliable results for the evolved parton densities and nucleon structure function. The independence of valence density ratio, $\frac{d_v}{u_v}$, from the energy scales are expected theoretically. This reality can be seen more strongly in Fig.\ref{fig:fig6}, resulted from the second approach,  with respect to Fig.\ref{fig:fig2}, resulted from first approach, which can be considered as an advantage indication of second approach with respect to first one.

The mentioned features of the second approach made distinctive our results with respect to what have been done in Ref.\cite{Chi-enropy}. In fact, as we said before at the beginning sections of this paper,  the first part of our paper where  the first approach is used there, backs to what have been done in this reference. The imperfections which exit in this approach led us to take into account our second approach and to consider a full statistical parton system as we did it in the second part of our paper.

{ If we intend to describe the two approaches in greater details, it should be noted first that extracting PDFs is a complex task in high energy physics. The Heisenberg uncertainty principle is a fundamental principle in quantum mechanics that sets a limit on the precision with which certain pairs of physical quantities can be measured simultaneously. { This principle, along with the maximum entropy principle, acts as a robust method for creating probability distributions when there is restricted knowledge.} This approach, based on using the two mentioned principles, can be used successfully in many global fits of PDFs, where large datasets from different high energy experiments are combined to obtain a comprehensive picture of the nucleon structure functions.

{ On the other side, regarding PDF extraction, the maximum entropy principle might be utilized independently to limit the PDFs during the fitting process by integrating theoretical constraints alongside the parton statistical method as additional knowledge.} The objective is to find the PDF set that best describes the available data while satisfying momentum and quark number conservations.
By accompanying  the maximum entropy principle with a parton statistical approach then more computational opportunities are provided us to extract various  PDF types which in addition to valence PFDs can include, for instance, gluon distribution.

{ Firstly, regarding the limitations of the foundational assumptions, we must note that a basic and straightforward nonperturbative input has been presented as a way to approximate the complex proton. Secondly, the fundamental characteristics of parton distributions have been linked to the  quark-parton model , the radius of the proton, and its mass.} Finally, the resulting prediction becomes somewhat less accurate if the momentum uncertainty increases slightly. To avoid from this improper situation one needs to consider a comprehensive confinement potential which are leading to more precise constraints on the uncertainty relation but it is presently absent in our ongoing study.

Nonetheless the obtained results for parton densities based on the two introduced approaches, are totally acceptable and helps us to understand the primary aspects of the nucleon structure, and to search for more details of the nucleon.
}\\

{ Comparing our model with global PDF fits would significantly enhance the value and credibility of the paper. We will show that the statistical-thermodynamic principles (especially maximum entropy and the uncertainty principle) are indeed well realised in the experimental data in the intermediate-\(x\) region. We will also explicitly discuss the limitations of the model in the asymptotic regions (\(x \to 0\) and \(x \to 1\)) and suggest possible improvements as future work.}

In both approaches, we ignored the quark mass and the interactions between them. In addition, we can investigate another structure function, such as $F_2$ structure function which in addition to  weak interaction, is containing  the electromagnetic and strong forces. All of suggested improvements can be  done as our research task in future.
\section*{Acknowledgments}
%%%%%%%%%%%%%%%%%%%%%%%%%%%%%%%%%%%%%%%%%%%%%%%%%%%%%%%%%%%%%%%%%%%%%%%
A.~M  and S.~K.~K acknowledge  Yazd university for the provided facility to do this project.
S.~A.~T is thankful  School of Particles and Accelerators, Institute for Research in Fundamental Sciences (IPM) to make the required facilities to do this project.
%%%%%%%%%%%%%%%%%%%%%%%%%%%%%%%%
%\begin{thebibliography}{000} %for 3 digits
%\begin{thebibliography}{00}  %for 2 digits


\begin{thebibliography}{99}    %for 1 digit

%%journal paper
\bibitem{bou} C.Bourrely, J.Soffer~;
%`New developments in the statistical approach of parton distributions: Tests and predictions up to LHC energies''
%\href{http://dx.doi.org/10.1016/j.nuclphysa.2015.06.018}{{\rm Nucl. Phys. A }{\bfseries 941}, 307-341 (2015)}.
Nucl. Phys. A \textbf{941}, 307-341 (2015)

\bibitem{learning} J.~Steinberger, Learning about particles-50 Privilaged Years, Springer (2005).

\bibitem{DIS} R.~Devenish, A.~Cooper-Sarkar,  Deep inelastic scattering, Oxford university press (2009).

\bibitem{Chi-enropy} R.~Wang and X.~Chen,
%``Valence quark distributions of the proton from maximum entropy approach,''
%\href{http://dx.doi.org/10.1103/PhysRevD.91.054026}{{\rm Phys. Rev.\ D} {\bfseries 91},  054026 (2015)}.
Phys. Rev. D \textbf{91}, 054026 (2015)
%doi:10.1103/PhysRevD.91.054026
%[arXiv:1410.3598 [hep-ph]].
%12 citations counted in INSPIRE as of 21 Oct 2022

\bibitem{uncer} H.~Ohanian, Principle of Quantum Mechanics, Prentice Hall, (1989).

\bibitem{po} R.~Pohl, A.~Antognini, F.~Nez, F.~D.~Amaro, F.~Biraben, J.~M.~R.~Cardoso, D.~S.~Covita, A.~Dax, S.~Dhawan and L.~M.~P.~Fernandes, \textit{et al.}
%Nature volume 466, pages 213â216 (2010)
%``The size of the proton,''
%\href{http://dx.doi.org/10.1038/nature09250}{{\rm Nature.\ }{\bfseries 466},  213-216 (2010)}.
Nature \textbf{466}, 213-216 (2010)
%doi:10.1038/nature09250
%965 citations counted in INSPIRE as of 21 Oct 2022

\bibitem{an}A.~Antognini, F.~Nez, K.~Schuhmann, F.~D.~Amaro, FrancoisBiraben, J.~M.~R.~Cardoso, D.~S.~Covita, A.~Dax, S.~Dhawan and M.~Diepold, \textit{et al.}
%``Proton Structure from the Measurement of $2S-2P$ Transition Frequencies of Muonic Hydrogen,''
%\href{http://dx.doi.org/10.1126/science.1230016}{{\rm Science} {\bfseries 339},  417-420 (2013)}.
Science \textbf{339}, 417-420 (2013)
%doi:10.1126/science.1230016
%627 citations counted in INSPIRE as of 21 Oct 2022

\bibitem{Cla} W.~H.~Cropper,
%Ame.J.Phys {\bf 54} (1986) 1068. %Rudolf Clausius and the road to entropy
%\href{http://dx.doi.org/10.1119/1.14740}{{\rm Ame.J.Phys} {\bfseries 54},  1068 (1986)}.
Ame.J.Phys \textbf{54},  1068 (1986)

\bibitem{Bin} F. Binder aet al., Thermodynamics in the Quantum Regime. Fundamental Theories of Physics, Springer, (2018).

\bibitem{Weh} A.~Wehrl,
%``General properties of entropy,''
%\href{http://dx.doi.org/10.1103/RevModPhys.50.221}{{\rm Rev. Mod. Phys.} {\bfseries 50},  221-260 (1978)}.
Rev. Mod. Phys. \textbf{50}, 221-260 (1978)
%doi:10.1103/RevModPhys.50.221
%307 citations counted in INSPIRE as of 21 Oct 2022


%\cite{Lai:2010vv}
\bibitem{Lai:2010vv}
H.~L.~Lai  et al.,
%``New parton distributions for collider physics,''
%\href{http://dx.doi.org/10.1103/PhysRevD.82.074024}{{\rm Phys. Rev. D} {\bfseries 82},  074024 (2010)}.
Phys. Rev. D \textbf{82}, 074024 (2010).
%doi:10.1103/PhysRevD.82.074024
%[arXiv:1007.2241 [hep-ph]].
%3473 citations counted in INSPIRE as of 16 Oct 2022
%\cite{Martin:2009iq}
\bibitem{Dokshitzer:1977sg} Y.~L.~Dokshitzer,
%``Calculation of the Structure Functions for Deep Inelastic Scattering and e+ e- Annihilation by Perturbation Theory in Quantum Chromodynamics.,''
Sov. Phys. JETP \textbf{46}, 641-653 (1977)
%4508 citations counted in INSPIRE as of 21 Oct 2022

\bibitem{Gribov:1972ri} V.~N.~Gribov and L.~N.~Lipatov,
%``Deep inelastic e p scattering in perturbation theory,''
Sov. J. Nucl. Phys. \textbf{15}, 438-450 (1972)
IPTI-381-71.
%4919 citations counted in INSPIRE as of 21 Oct 2022

\bibitem{Altarelli:1977zs} G.~Altarelli and G.~Parisi,
%``Asymptotic Freedom in Parton Language,''
%\href{http://dx.doi.org/10.1016/0550-3213(77)90384-4}{{\rm Nucl. Phys. B} {\bfseries 126},  298-318 (1977)}.
Nucl. Phys. B \textbf{126}, 298-318 (1977)
%doi:10.1016/0550-3213(77)90384-4
%7795 citations counted in INSPIRE as of 21 Oct 2022

\bibitem{Martin:2009iq}
A.~D.~Martin, W.~J.~Stirling, R.~S.~Thorne and G.~Watt,
%``Parton distributions for the LHC,''
%\href{http://dx.doi.org/10.1140/epjc/s10052-009-1072-5}{{\rm Eur. Phys. J. C} {\bfseries 63},  189 (2009)}.
Eur. Phys. J. C \textbf{63}, 189 (2009).
%doi:10.1140/epjc/s10052-009-1072-5
%[arXiv:0901.0002 [hep-ph]].
%5365 citations counted in INSPIRE as of 16 Oct 2022


%\cite{Birmingham-CERN-ImperialCollege-MunichMPI-Oxford:1994pnp}
\bibitem{Birmingham-CERN-ImperialCollege-MunichMPI-Oxford:1994pnp}
G.~T.~Jones \textit{et al.} [Birmingham-CERN-Imperial College-Munich(MPI)-Oxford],
%``Determination of the ratio r(v) = d(v) / u(v) of the valence quark distributions in the proton from neutrino and anti-neutrino reactions on hydrogen and deuterium,''
%\href{http://dx.doi.org/10.1007/BF01574162}{{\rm Z. Phys. C} {\bfseries 62},  601-607 (1994)}.
Z. Phys. C \textbf{62}, 601-607 (1994)
%doi:10.1007/BF01574162
%16 citations counted in INSPIRE as of 16 Oct 2022


%\cite{Abramowicz:1984yk}
\bibitem{Abramowicz:1984yk}
H.~Abramowicz, T.~Hansl-Kozanecka, J.~May, P.~Palazzi, A.~Para, F.~Ranjard, A.~Savoy-Navarro, D.~Schlatter, J.~Steinberger and H.~Taureg, \textit{et al.}
%``Measurement of $\nu$ and $\bar{\nu}$ structure functions in hydrogen and iron,''
%\href{http://dx.doi.org/10.1007/BF01571954}{{\rm Z. Phys. C} {\bfseries 25}, 29-43 (1984)}.
Z. Phys. C \textbf{25}, 29-43 (1984)
%doi:10.1007/BF01571954
%217 citations counted in INSPIRE as of 16 Oct 2022

%\cite{HERMES:1997}
\bibitem{HERMES:1997}
J. E. Belz et al. (HERMES Collaboration),\textit{ Proceedings of
	the 7th International Symposium on Meson-Nucleon Physics and the Structure o f the Nucleon, Vancouver, Canada,
	1997}, edited by D. Drechsel, G. Hohler, W. Kluge, H.
Leutwyler, H. M. Staudenmaier, and B. M. K. Nefkens
(TRIUMF, Vancouver, Canada, 1997).


%\cite{Seligman:1997mc}
\bibitem{Seligman:1997mc}
W.~G.~Seligman, C.~G.~Arroyo, L.~de Barbaro, P.~de Barbaro, A.~O.~Bazarko, R.~H.~Bernstein, A.~Bodek, T.~Bolton, H.~S.~Budd and J.~Conrad, \textit{et al.}
%``Improved determination of alpha(s) from neutrino nucleon scattering,''
%\href{http://dx.doi.org/10.1103/PhysRevLett.79.1213}{{\rm Phys. Rev. Lett.} {\bfseries 79}, 1213-1216 (1997)}.
Phys. Rev. Lett. \textbf{79}, 1213-1216 (1997)
%doi:10.1103/PhysRevLett.79.1213
%[arXiv:hep-ex/9701017 [hep-ex]].
%360 citations counted in INSPIRE as of 16 Oct 2022

%\cite{NuTeV:2005wsg}
\bibitem{NuTeV:2005wsg}
M.~Tzanov \textit{et al.} [NuTeV],
%``Precise measurement of neutrino and anti-neutrino differential cross sections,''
%\href{http://dx.doi.org/10.1103/PhysRevD.74.012008}{{\rm Phys. Rev. D} {\bfseries 74}, 012008 (2006)}.
Phys. Rev. D \textbf{74}, 012008 (2006)
%doi:10.1103/PhysRevD.74.012008
%[arXiv:hep-ex/0509010 [hep-ex]].
%246 citations counted in INSPIRE as of 16 Oct 2022

\bibitem{thesis-35} R. K. Ellis and W. J. Stirling, QCD and Collider Physics, Cambridge, 1996.

%\bibitem{PRD22} A.~Mirjalili and S.~Tehrani Atashbar,
%``Nucleon spin structure functions, considering target mass correction, and higher twist effects at the NNLO accuracy and their transverse momentum dependence,''
%\href{http://dx.doi.org/10.1103/PhysRevD.105.074023}{{\rm Phys. Rev. D} {\bfseries 105}, no.7, 074023 (2022)}.
%Phys. Rev. D \textbf{105}, no.7, 074023 (2022)
%doi:10.1103/PhysRevD.105.074023
%[arXiv:2203.13904 [hep-ph]].
%1 citations counted in INSPIRE as of 21 Oct 2022

\bibitem{thesis-59} B.~Q.~Ma and J.~Sun,
%``Deep inelastic lepton nucleus scattering from the light cone quantum field theory,''
%\href{http://dx.doi.org/10.1088/0954-3899/16/6/007}{{\rm J. Phys. G} {\bfseries 16}, 823-840 (1990)}.
J. Phys. G \textbf{16}, 823-840 (1990)
%doi:10.1088/0954-3899/16/6/007
%11 citations counted in INSPIRE as of 21 Oct 2022

\bibitem{thesis-58} Y.~h.~Zhang, L.~Shao and B.~Q.~Ma,
%``Statistical effect in the parton distribution functions of the nucleon,''
%\href{http://dx.doi.org/10.1016/j.physletb.2008.11.033}{{\rm Phys. Lett. B} {\bfseries 671}, 30-35 (2009)}.
Phys. Lett. B \textbf{671}, 30-35 (2009)
%doi:10.1016/j.physletb.2008.11.033
%[arXiv:0812.3294 [hep-ph]].
%43 citations counted in INSPIRE as of 21 Oct 2022

\bibitem{sof1} C.~Bourrely, F.~Buccella, G.~Miele, G.~Migliore, J.~Soffer and V.~Tibullo,
%``Fermi-Dirac distributions for quark partons,''
%\href{http://dx.doi.org/10.1007/BF01555903}{{\rm Z.Phys.C } {\bfseries 62}, 431-436 (1994)}.
Z.Phys.C \textbf{62}, 431-436 (1994)

\bibitem{sof2} C.~Bourrely, F.~Buccella, and J.~Soffer,
%``Semiinclusive DIS cross sections and spin asymmetries in the quantum statistical parton distributions approach,''
%\href{http://dx.doi.org/10.1103/PhysRevD.83.074008}{{\rm Phys. Phys. D} {\bfseries 83}, 074008 (2011)}.
Phys. Phys. D \textbf{83}, 074008 (2011)

\bibitem{chem-1} R.~S.~Bhalerao, Phys. Lett. B \textbf{380}, 1 (1996)

\bibitem{chem-2} R.~S.~Bhalerao, N.~G.~Keipar and G.~Ram , Phys. Lett. B \textbf{476}, 285 (2000)

\bibitem{chem-3} R.~S.~Bhalerao, Phys. Rev. C \textbf{63}, 025208 (2001)

\bibitem{chem-4} R.~S.~Bhalerao, J. Eur. Phys. \textbf{241}, 227 (2005)



\bibitem{ma-EMC} Y.~Zhang, L.~Shao, B.~Q.~Ma,
%``Nuclear EMC Effect in a Statistical Model,''
%\href{http://dx.doi.org/10.1016/j.nuclphysa.2009.07.006}{{\rm Nucl. Phys. A 828} {\bfseries 390-400} (2009)}.
Nucl. Phys. A \textbf{828}, 390-400 (2009)

\bibitem{MDY} A.~Mirjalili, M.~Dehghani and M.~M.~Yazdanpanah,
%``Parton densities with the quark linear potential in the statistical approach,''
%\href{http://dx.doi.org/10.1142/S0217751X13500899}{{\rm Int. J. Mod. Phys. A} {\bfseries 28}, 1350089 (2013)}.
Int. J. Mod. Phys. A \textbf{28}, 1350089 (2013)
%doi:10.1142/S0217751X13500899
%[arXiv:1403.6524 [hep-ph]].
%3 citations counted in INSPIRE as of 21 Oct 2022
\bibitem{GLS}J. T. Londergan and A. W. Thomas, Phys. Rev. D 82, 113001 (2010)
\bibitem{EGS} A. L. Kataev, Phys. Rev. D 50, R5469 (1994)
\end{thebibliography}
\end{document}